\documentclass[acus]{pac99}
\usepackage{epsfig}

\begin{document}
\title{NONLINEAR ACCELERATOR PROBLEMS VIA WAVELETS:\\
7. INVARIANT CALCULATIONS IN HAMILTONIAN PROBLEMS}
\author{A.~Fedorova, M.~Zeitlin, IPME, RAS, St.~Petersburg, Russia
  \thanks{ e-mail: zeitlin@math.ipme.ru}
  \thanks{http://www.ipme.ru/zeitlin.html;
   http://www.ipme.nw.ru/zeitlin.html}}
\maketitle
\begin{abstract}
In this series of eight papers  we
present the applications of methods from
wavelet analysis to polynomial approximations for
a number of accelerator physics problems.
In this paper we consider invariant formulation of nonlinear
(Lagrangian or Hamiltonian) dynamics on semidirect structure
(relativity or dynamical groups) and corresponding invariant
calculations via CWT.
\end{abstract}

\section{INTRODUCTION}
This is the seventh part of our eight presentations in which we consider
applications of methods from wavelet analysis to nonlinear accelerator
physics problems.
 This is a continuation of our results from [1]-[8],
in which we considered the applications of a number of analytical methods from
nonlinear (local) Fourier analysis, or wavelet analysis, to nonlinear
accelerator physics problems
 both general and with additional structures (Hamiltonian, symplectic
or quasicomplex), chaotic, quasiclassical, quantum. Wavelet analysis is
a relatively novel set of mathematical methods, which gives us a possibility
to work with well-localized bases in functional spaces and with the
general type of operators (differential, integral, pseudodifferential) in
such bases.
In contrast  with parts 1--4 in parts 5--8 we try to take into account
before using power analytical approaches underlying algebraical, geometrical,
topological structures related to kinematical, dynamical and hidden
symmetry of physical problems.
We described  a number of concrete problems in parts 1--4.
The most interesting case is the dynamics of spin-orbital motion (part 4).
In section 2 we consider dynamical consequences of covariance properties
regarding to relativity (kinematical) groups and continuous wavelet transform
 (CWT) (in section 3)
as a method for the solution of dynamical problems.
We introduce the semidirect product structure, which allows us to
consider from general point of view all relativity groups such as Euclidean,
Galilei, Poincare. Then  we consider the Lie-Poisson equations and
obtain the manifestation of semiproduct structure of (kinematic) symmetry group
on dynamical level. So, correct description of dynamics is a consequence of
correct understanding of real symmetry of the concrete problem.
We consider the Lagrangian theory related to semiproduct structure
and explicit form of variation principle and corresponding (semidirect)
Euler-Poincare equations.
In section 3 we consider CWT and the corresponding
analytical technique which allows to consider covariant wavelet analysis.
In  part 8 we consider in the particular case of affine Galilei group
with the semiproduct structure  the
corresponding orbit technique for constructing different types of invariant
wavelet bases.

\section{Dynamics on Semidirect Products}
Relativity groups such as Euclidean, Galilei or Poincare groups  are the
particular cases of semidirect product construction, which is very useful and
simple general construction in the group theory [9]. We may consider as a basic
example the Euclidean group $SE(3)=SO(3)\bowtie{\bf R}^3$, the semidirect
product of rotations and translations. In general case we have $S=G\bowtie V$,
where group G (Lie group or automorphisms group) acts on a vector space V and
on its dual $V^*$.
Let $V$ be a vector space and $G$ is the Lie group, which acts on the left by
linear maps on V (G also acts on the left on its dual space $V^*$).
The Lie algebra of S is the semidirect product  Lie algebra,
$s={\mathcal{G}} \bowtie V$ with brackets
$
[(\xi_1,v_1),(\xi_2,v_2)]=([\xi_1,\xi_2],\xi_1v_2-\xi_2v_1),
$
where the induced action of $\mathcal{G}$ by concatenation is denoted
as $\xi_1 v_2$.
Let $(g,v)\in S=G\times V, \quad (\xi,u)\in
s={\mathcal{G}}\times V$, $(\mu,a)\in s^*={\mathcal G}^*\times V^*$, $g\xi=Ad_g\xi$,
$g\mu=Ad^*_{g^{-1}}\mu$, $ga$ denote the induced left action of $g$ on $a$ (the
left action of G on V induces a left action on $V^*$ --- the inverse of the
transpose of the action on V), $\rho_v: {\cal G}\to V$ is a linear map
given by $\rho_v(\xi)=\xi v$, $\rho^*_v: V^*\to{\cal G}^*$ is its dual.
Then adjoint and coadjoint actions are given by simple concatenation:
$(g,v)(\xi,u)=(g\xi,gu-(g\xi)v)$,
$(g,v)(\mu,a)=(g\mu+\rho^*_v(ga),ga)$.
Also, let be $\rho^*_v a=v\diamond a\in{\cal G^*}$ for
$a\in V^*$, which is a bilinear operation in $v$ and $a$. So, we have the coadjoint
action:
$
(g,v)(\mu,a)=(g\mu+v\diamond(ga),ga).
$
Using concatenation notation for Lie algebra actions we have alternative
definition of $v\diamond a\in{\mathcal G}^*$.
For all $v\in V$, $a\in V^*$, $\eta\in{\mathcal G}$ we have
$
<\eta a,v>=-<v\diamond a, \eta>
$.

Now we consider the manifestation of semiproduct structure
of symmetry group on dynamical level.
Let $F,G$ be real valued functions on the dual space ${\mathcal G}^*$,
 $\mu\in{\mathcal G}^*$. Functional
derivative of F at $\mu$ is the unique element $\delta F/\delta\mu\in{\mathcal
G}$:
$
\lim_{\epsilon\to 0}
\lbrack F(\mu+\epsilon\delta\mu)-F(\mu)\rbrack/\epsilon=
<\delta\mu,{\delta F}/{\delta\mu}>
$
for all $\delta\mu\in{\mathcal G}^*$, $<,>$ is pairing between $\mathcal G^*$ and
$\mathcal G$.
Define the $(\pm)$ Lie-Poisson brackets by
$\{F,G\}_\pm(\mu)=\pm <\mu,\lbrack{\delta F}/{\delta\mu},
{\delta G}/{\delta\mu}\rbrack>.$
The Lie-Poisson equations, determined by
$\dot{F}=\{F,H\}$
or intrinsically
$\dot{\mu}=\mp ad^*_{\partial H/\partial\mu}\mu.$
For the left representation of G on V $\pm$ Lie-Poisson bracket of two
functions $f,k: s^*\to {\bf R}$ is given by
\begin{eqnarray}
&&\{f,k\}_{\pm}(\mu, a)=\pm <\mu,\lbrack\frac{\delta f}{\delta\mu},
\frac{\delta k}{\delta\mu}\rbrack>\\
&&\pm
<a,\frac{\delta f}{\delta\mu}\frac{\delta k}{\delta a}-
\frac{\delta k}{\delta\mu}\frac{\delta f}{\delta a}>,\nonumber
\end{eqnarray}
where $\delta f/\delta\mu\in{\mathcal G}$, $\delta f/\delta a\in V$ are
the functional derivatives of f. The Hamiltonian vector field of
$h: s^*\in{\bf R}$ has the expression
$X_h(\mu,a)=\mp(ad^*_{\delta h/\delta\mu}\mu-{\delta h}/{\delta
a}\diamond a, -{\delta h}/{\delta\mu}a).$
Thus, Hamiltonian equations on the dual of a semidirect product are [9]:
\begin{eqnarray}\label{eq:mua}
\dot{\mu}=\mp ad^*_{\delta h / \delta\mu}\mu\pm\frac{\delta h}{\delta a}\diamond
 a,\quad
\dot{a}=\pm\frac{\delta h}{\delta\mu} a
\end{eqnarray}
So, we can see the explicit contribution to the Poisson brackets
and the equations of motion
which come from the semiproduct structure.

Now we consider according to [9] Lagrangian side of a theory.
This approach is based on variational principles with symmetry and is not
dependent on Hamiltonian formulation, although it is demonstrated in [9] that this
purely Lagrangian formulation is equivalent to the Hamiltonian formulation
on duals of semidirect product (the
corresponding Legendre transformation is a diffeomorphism).
We consider the case of the left representation and
the left invariant  Lagrangians ($\ell$ and L),
which depend in additional on another parameter $a\in V^*$ (dynamical
parameter),
where V is  representation space for the Lie group G and L has  an invariance
property related to both arguments. It should be noted that the resulting
equations of motion, the Euler-Poincare equations, are not the Euler-Poincare
equations for the semidirect product Lie algebra ${\mathcal G}\bowtie V^*$ or
${\mathcal G}\bowtie V$.
So, we have the following:

{\bf 1.} There is a left representation of Lie group G on the vector space V and G
acts in the natural way on the left on $TG\times V^*: h(v_g,a)=(hv_g,ha)$.
{\bf 2.} The function $L: TG\times V^*\in{\bf R}$ is the left G-invariant.
{\bf 3.} Let $a_0\in V^*$, Lagrangian $L_{a_0}: TG\to{\bf R}$,
$L_{a_0}(v_g)=L(v_g,a_0)$. $L_{a_0}$ is left invariant under the lift to TG of
the left action of $G_{a_0}$ on G, where $G_{a_0}$ is the isotropy group of
$a_0$.
{\bf 4.} Left G-invariance of L permits us to define
$\ell:{\mathcal G}\times V^*\to{\bf R}$
by
$\ell(g^{-1}v_g,g^{-1}a_0)=L(v_g,a_0).$
This relation defines for any $\ell:{\mathcal G}\times V^*\to{\bf R}$ the left
G-invariant function $L: TG\times V^*\to{\bf R}$.
{\bf 5.} For a curve $g(t)\in G$ let be
$\xi(t):=g(t)^{-1}\dot{g}(t)$
and define the curve $a(t)$ as the unique solution of the following linear
differential equation with time dependent coefficients
$\dot{a}(t)=-\xi(t)a(t),$
with initial condition $a(0)=a_0$. The solution can be written as
$a(t)=g(t)^{-1}a_0$.

Then we have four equivalent descriptions of the corresponding dynamics:
{\bf 1.} If $a_0$ is fixed then Hamilton's variational principle
$\delta\int_{t_1}^{t_2}L_{a_0}(g(t),\dot{g}(t){\rm d}t=0$
holds for variations $\delta g(t)$ of $g(t)$ vanishing at the endpoints.
{\bf 2.} $g(t)$ satisfies the Euler-Lagrange equations for $L_{a_0}$ on G.
{\bf 3.} The constrained variational principle
$\delta\int_{t_1}^{t_2}\ell(\xi(t),a(t)){\rm d}t=0$
holds on ${\mathcal G}\times V^*$, using variations of $\xi$ and $a$ of the form
 $\delta\xi=\dot{\eta}+[\xi,\eta]$, $\delta a=-\eta a$, where
$\eta(t)\in{\mathcal G}$ vanishes at the endpoints.
{\bf 4.} The Euler-Poincare equations hold on ${\mathcal G}\times V^*$
\begin{equation}
\frac{{\rm d}}{{\rm d}t}\frac{\delta\ell}{\delta\xi}=
ad_\xi^*\frac{\delta\ell}{\delta\xi}+\frac{\delta\ell}{\delta a}\diamond a
\end{equation}
So, we may apply our wavelet methods either on the level of variational formulation
or on the level of Euler-Poincare equations.

\section{Continuous Wavelet Transform}
  Now we need take into account the Hamiltonian
or Lagrangian structures related with systems (2) or  (3).
Therefore, we need to consider generalized wavelets, which
allow us to consider the corresponding  structures  instead of
compactly supported wavelet representation from parts 1-4.
In wavelet analysis the following three concepts are used now:
1).\ a square integrable representation $U$ of a group $G$,
2).\ coherent states (CS) over G,
3).\ the wavelet transform associated to U. We consider now
their unification [10]-[12].
Let $G$ be a locally compact group and $U_a$  strongly continuous,
irreducible, unitary representation of G on Hilbert space ${\mathcal H}$.
Let $H$ be a closed subgroup of $G$, $X=G/H$ with (quasi) invariant measure $\nu$
and $\sigma: X=G/H\to G$ is a Borel section in a principal bundle $G\to G/H$.
Then we say that $U$ is square integrable $mod(H,\sigma)$ if there exists a
non-zero vector $\eta\in{\mathcal H}$ such that
$0<\int_X|<U(\sigma(x))\eta|\Phi>|^2{\rm d}\nu(x)=<\Phi|A_\sigma\Phi>\
<\infty,$
$ \forall\Phi\in{\mathcal H}$.
Given such a vector $\eta\in{\mathcal H}$ called admissible for $(U,\sigma)$ we
define the family of (covariant) coherent states or wavelets, indexed by points
$x\in X$, as the orbit of $\eta$ under $G$, though the representation $U$ and the
section $\sigma$ [10]-[12]:
$S_\sigma={\eta_{\sigma(x)}=U(\sigma(x))\eta|x\in X}$.
So, coherent states or wavelets are simply the elements of the orbit under U of
a fixed vector $\eta$ in representation space.
We have the following fundamental properties:
{\bf 1.}Overcompleteness:
 the set $S_\sigma$ is total in ${\mathcal H}:(S_\sigma)^\perp={0}$.
{\bf 2.} Resolution property:
the square integrability condition  may be represented as a
resolution relation:
$\int_X|\eta_\sigma(x)><\eta_{\sigma(x)}|{\rm d}\nu(x)=A_\sigma,$
where $A_\sigma$ is a bounded, positive operator with a densely defined
inverse. Define the linear map
$W_\eta: {\mathcal H}\to L^2(X,{\rm
d}\nu),(W_\eta\Phi)(x)=<\eta_{\sigma(x)}|\Phi>.$
Then the range $H_\eta$ of $W_\eta$ is complete with respect to the scalar
product $<\Phi|\Psi>_\eta=<\Phi|W_\eta A^{-1}_\sigma W^{-1}_\eta\Psi>$ and $W_\eta$ is
unitary operator from ${\mathcal H}$ onto ${\mathcal H}_\eta$.
$W_\eta$ is Continuous Wavelet Transform (CWT).
{\bf 3.} Reproducing kernel.
The orthogonal projection from $L^2(X,{\rm d}\nu)$ onto ${\mathcal H}_\eta$ is
an integral operator $K_\sigma$ and $H_\eta$ is a reproducing kernel Hilbert
space of functions:
$\Phi(x)=\int_XK_\sigma(x,y)\Phi(y){\rm d}\nu(y), \quad \forall \Phi\in{\mathcal
H}_\eta.$
The kernel is given explicitly by
$K_\sigma(x,y)=<\eta_{\sigma(x)}A_\sigma^{-1}\eta_{\sigma(y)}>$, if
$\eta_{\sigma(y)}\in D(A^{-1}_\sigma)$, $\forall y\in X$.
So, the function $\Phi\in L^2(X,{\rm d}\nu)$ is a wavelet transform (WT)
iff it satisfies this reproducing relation.
{\bf 4.} Reconstruction formula.
The WT $W_\eta$ may be inverted on its range by the adjoint operator,
$W_\eta^{-1}=W_\eta^*$ on ${\mathcal H}_\eta$ to obtain for
$\eta_{\sigma(x)}\in D(A_\sigma^{-1})$, $\forall x\in X$
$W_\eta^{-1}\Phi=\int_X\Phi(x)A_\sigma^{-1}\eta_{\sigma(x)}{\rm d}\nu(x), \
\Phi\in{\mathcal H}_\eta.$
This is inverse WT.
If $A_\sigma^{-1}$ is bounded then $S_\sigma$ is called a frame, if
$A_\sigma=\lambda I$ then $S_\sigma$ is called a tight frame. This two cases
are generalization of a simple case, when $S_\sigma$ is an (ortho)basis.

The most simple cases of this construction are:\\
{\bf 1.} $H=\{e\}$. This is the standard construction of WT over a locally compact
group. It should be noted that the square integrability of U is equivalent to
U belonging to the discrete series. The most simple example is related to
 the affine $(ax+b)$ group and yields the usual one-dimensional wavelet
analysis
$[\pi(b,a)f](x)=\frac{1}{\sqrt{a}}f\left(\frac{x-b}{a}\right).$
For $G=SIM(2)={\bf R}^2\bowtie({\bf R}^{+}_*\times SO(2))$,
the similitude group of the plane, we have the corresponding two-dimensional
wavelets.
{\bf 2.} $H=H_\eta$, the isotropy (up to a phase) subgroup of $\eta$: this is the
case of the Gilmore-Perelomov CS. Some cases of group G are:
{\bf a).} Semisimple groups, such as SU(N), SU(N$|$M), SU(p,q), Sp(N,{\bf R}).
{\bf b).} the Weyl-Heisenberg  group $G_{WH}$ which leads to the Gabor
functions, i.e. canonical (oscillator)coherent states associated
with windowed Fourier
transform or Gabor transform (see also part 6):
$[\pi(q,p,\varphi)f](x)=\exp(i\mu(\varphi-p(x-q))f(x-q)$.
In this case H is the  center of $G_{WH}$.
In both cases time-frequency plane corresponds to the phase
space of group representation.
{\bf c).} The similitude group SIM(n) of ${\bf R}^n(n\ge3)$:
for $H=SO(n-1)$ we have
the axisymmetric n-dimensional wavelets.
{\bf d).} Also we have
 the case of bigger group, containing
both affine and We\-yl-\-Hei\-sen\-berg group, which interpolate between
affine wavelet analysis and windowed Fourier analysis: affine
Weyl--Heisenberg group [12].
{\bf e).} Relativity groups. In a nonrelativistic setup, the natural kinematical
group is the (extended) Galilei group. Also we may adds independent space and
time dilations and obtain affine Galilei group. If we restrict the dilations by
the relation $a_0=a^2$, where $a_0, a$ are the time and space dilation we
obtain the Galilei-Schr\"odinger group, invariance group of both Schr\"odinger
and heat equations. We consider these examples in the next section. In the same
way we may consider as kinematical group the Poincare group. When $a_0=a$ we
have affine Poincare or Weyl-Poincare group. Some useful generalization of that
affinization construction we consider for the case of hidden metaplectic structure
 in part 6.
But the usual representation  is not
square--integrable and must be modified: restriction of the
representation to a suitable quotient space of the group (the
associated phase space in our case) restores square --
integrability: $G\longrightarrow$ homogeneous space.
Our goal is applications of these results to problems of
Hamiltonian dynamics and as consequence we need to take into account
symplectic nature of our dynamical problem.
 Also, the symplectic and wavelet structures
 must be consistent (this must
be resemble the symplectic or Lie-Poisson integrator theory).
 We use the
point of view of geometric quantization theory (orbit method)
instead of harmonic analysis. Because of this we can consider
(a) -- (e) analogously. In next part we consider construction of invariant
bases.

We are very grateful to M.~Cornacchia (SLAC),
W.~Her\-r\-man\-nsfeldt (SLAC)
Mrs. J.~Kono (LBL) and
M.~Laraneta (UCLA) for
 their permanent encouragement.

\end{document}